\documentclass[aps,prl,twocolumn,amsmath,amssymb,nofootinbib,superscriptaddress]{revtex4-1}
\usepackage{times}
\usepackage[pdftex]{graphicx}
\usepackage{dcolumn}
\usepackage{bm}
\usepackage{amsmath}
\usepackage{indentfirst}
\usepackage{float}
\usepackage[colorlinks]{hyperref}
\usepackage[dvipsnames]{xcolor}

\usepackage{makecell}
\usepackage{diagbox}

\newcommand{\gcc}[1]{{\color{black}{#1}}}

\newcommand{\oo}{\ddot{\text{o}}}
\newcommand{\fxeb}{\mathcal{F}_{{\rm XEB}}}

\begin{document}

\title{Redefining the Quantum Supremacy Baseline With a New Generation Sunway Supercomputer}



\author{Xin Liu}
\affiliation{National Supercomputing Center in Wuxi, 
Wuxi, 
Jiangsu, China}

\author{Chu Guo}
\email{guochu604b@gmail.com}
\affiliation{Henan Key Laboratory of Quantum Information and Cryptography, Zhengzhou,
Henan 450000, China}

\author{Yong Liu}
\email{alexander\_liu\_99@163.com}
\affiliation{National Supercomputing Center in Wuxi, 
Wuxi, 
Jiangsu, China}

\author{Yuling Yang}
\affiliation{National Supercomputing Center in Wuxi, 
Wuxi, 
Jiangsu, China}

\author{Jiawei Song}
\affiliation{National Supercomputing Center in Wuxi, 
Wuxi, 
Jiangsu, China}

\author{Jie Gao}
\affiliation{National Supercomputing Center in Wuxi, 
Wuxi, 
Jiangsu, China}

\author{Zhen Wang}
\affiliation{National Supercomputing Center in Wuxi, 
Wuxi, 
Jiangsu, China}

\author{Wenzhao Wu}
\affiliation{National Supercomputing Center in Wuxi, 
Wuxi, 
Jiangsu, China}

\author{Dajia Peng}
\affiliation{Department of Earth System Science, Ministry of Education Key Laboratory for Earth System Modeling, Institute for Global Change Studies,Tsinghua University, Beijing 100084, China}

\author{Pengpeng Zhao}
\affiliation{National Supercomputing Center in Wuxi, 
Wuxi, 
Jiangsu, China}

\author{Fang Li}
\email{38349735@qq.com}
\affiliation{National Supercomputing Center in Wuxi, 
Wuxi, 
Jiangsu, China}

\author{He-Liang Huang}
\affiliation{Henan Key Laboratory of Quantum Information and Cryptography, Zhengzhou,
Henan 450000, China}
\affiliation{Shanghai Branch, CAS Centre for Excellence and Synergetic Innovation Centre in Quantum Information and Quantum Physics,\\
University of Science and Technology of China, Hefei, Anhui 201315, China}

\author{Haohuan Fu}
\email{haohuan@tsinghua.edu.cn}
\affiliation{Department of Earth System Science, Ministry of Education Key Laboratory for Earth System Modeling, Institute for Global Change Studies,Tsinghua University, Beijing 100084, China}

\author{Dexun Chen}
\affiliation{National Supercomputing Center in Wuxi, 
Wuxi, 
Jiangsu, China}

\begin{abstract}
A major milestone in the era of noisy intermediate scale quantum computers is \textit{quantum supremacy} [Nature \textbf{574}, 505 (2019)] claimed on the Sycamore quantum processor of $53$ qubits, 
which can perform a random circuit sampling task within $200$ seconds while the same task is estimated to require a runtime of $10,000$ years on Summit.
This record has been renewed with two recent experiments on the Zuchongzhi $2.0$ ($56$ qubits) and Zuchongzhi $2.1$ ($60$ qubits) quantum processors. On the other front of quantum supremacy comparison, there has also been continuous improvements on both the classical simulation algorithm as well as the underlying hardware. And a fair justification of the computational advantages for those quantum supremacy experiments would require to practically simulate the same problems on current top supercomputers, which is still in lack.
Here we report the full-scale simulations of these problems on new generation Sunway supercomputer, based on a customized tensor network contraction algorithm. 
Our benchmark shows that the most challenging sampling task performed on Sycamore can be accomplished within $1$ week, thus collapsing the quantum supremacy claim of Sycamore. 
Additionally, we show that the XEB fidelities of the \textit{quantum supremacy circuits} with up to $14$ cycles can be verified in minutes, which also provides strong consistency check for quantum supremacy experiments. 
Our results redefine quantum supremacy baseline using the new generation Sunway supercomputer.


\end{abstract}

\date{\today}
\maketitle

\address{}

\vspace{8mm}

\section{Introduction}
Ever since initially proposed in 1982~\cite{Feynman1982}, quantum computers have long held the belief to be able to efficiently solve certain computational problems that are intractable for classical computers. After $40$ years of theoretical and experimental developments~\cite{Shor1994,krantz2019quantum,huang2020superconducting,slussarenko2019photonic,blatt2012quantum,bruzewicz2019trapped,biamonte2017quantum,mcardle2020quantum}, it has now come to the era of noisy intermediate scale quantum computers which can manipulate several tens of noisy physical qubits~\cite{Preskill2018}. A major experimental milestone achieved along this way is the \textit{quantum supremacy} experiment conducted with the $53$-qubit superconducting quantum processor, entitled as Sycamore, by Google in 2019~\cite{AruteMartinisQuantumSupremacy2019}, which demonstrates that for the specific task of sampling from a random quantum circuit, Sycamore can be $10^9$ times faster than the best classical supercomputer Summit. Recently this record has been renewed with the $56$-qubit and $60$-qubit quantum processors entitled as Zuchongzhi $2.0$ and Zuchongzhi $2.1$ respectively~\cite{WuPan2021,ZhuPan2021}. The rapid evolution of quantum processors also enables quantum algorithms on a larger scale~\cite{HavlivcekJay2019,Google2020,McardleYuan2020,HarriganBabbush2021,SaggioWalther2021,Google2021}.


Quantum supremacy is a competition between the best quantum and classical computers and it is a continuous instead of a single-shot effort. Pushing this competition to its extreme would be beneficial for both fields. In fact, both the classical simulation algorithms as well as the classical computing hardware have been upgrading rapidly along with the development of quantum computers. On the classical simulation algorithm side, a work from Alibaba in 2020 proposes a slicing and subtree reconfiguration scheme to search for near optimal tensor contract orders under a fixed memory bound~\cite{HuangChen2021}. When used in combination with tensor network contraction (TNC) algorithm, they estimate that the most difficult sampling task on Sycamore, namely sampling from a random quantum circuit of $20$ cycles (referred as Sycamore-$20$ afterwards), would only cost $19$ days on Summit~\cite{HuangChen2021}. 
Instead of simulating exactly the same sampling task as performed on Sycamore, Ref.~\cite{PanZhang2021} shows that $2$ million correlated exact amplitudes for Sycamore-$20$ can be computed with a $60$-GPU cluster in $5$ days, based on a customized big-head TNC algorithm. A similar strategy is also adapted and implemented on the new Sunway supercomputer with the runtime shortened to $304$ seconds~\cite{LiuChen2021}.
On the classical computing hardware side, highly efficient accelerators such as GPU and TPU have been upgrading rapidly and exascale computing systems are emerging. As an example, the new-generation NVIDIA A100 GPU has a $19.5$-TFLOPS single-precision performance and a $312$-TFLOPS half-precision performance.

At the time of writing, a full-scale simulation of the most difficult sampling tasks performed on those quantum processors, 
which integrates both novelties on the classical side, namely a state of the art TNC algorithm and a top supercomputer in the world,
has not been reported. In this work, we report a highly efficient and full-scale implementation of a customized TNC algorithm on the new generation Sunway supercomputer. We demonstrate that the runtime to generate a perfect sample for Sycamore-$20$ is $440$ seconds with single-precision arithmetic and $276$ seconds with mixed-precision arithmetic. As a result, the most difficult sampling task performed on Sycamore, namely sampling $1$ million bitstrings with $0.2\%$ fidelity can be accomplished in $1$ week, collapsing quantum supremacy claimed for Sycamore. 
Our results thus provide the \textit{quantum supremacy baseline} for future benchmark. 

Additionally, we verify three quantum supremacy circuits (defined as random quantum circuits with $12$ cycles or above), namely Sycamore-$12$, Sycamore-$14$ and Zuchongzhi $2.0$-$12$, by computing the exact amplitudes for $1$ million experimentally generated bitstrings and then computing the cross benchmarking (XEB) fidelities for those circuits. The obtained values for XEB fidelities are slightly lower, but within errorbar, than those values estimated from the simplified variants used in the quantum supremacy experiments. 
Our result makes possible the real-time verification of large-scale quantum supremacy circuits which could become an indispensable tool for future development of quantum processors.

\begin{figure*}
\includegraphics[width=2\columnwidth]{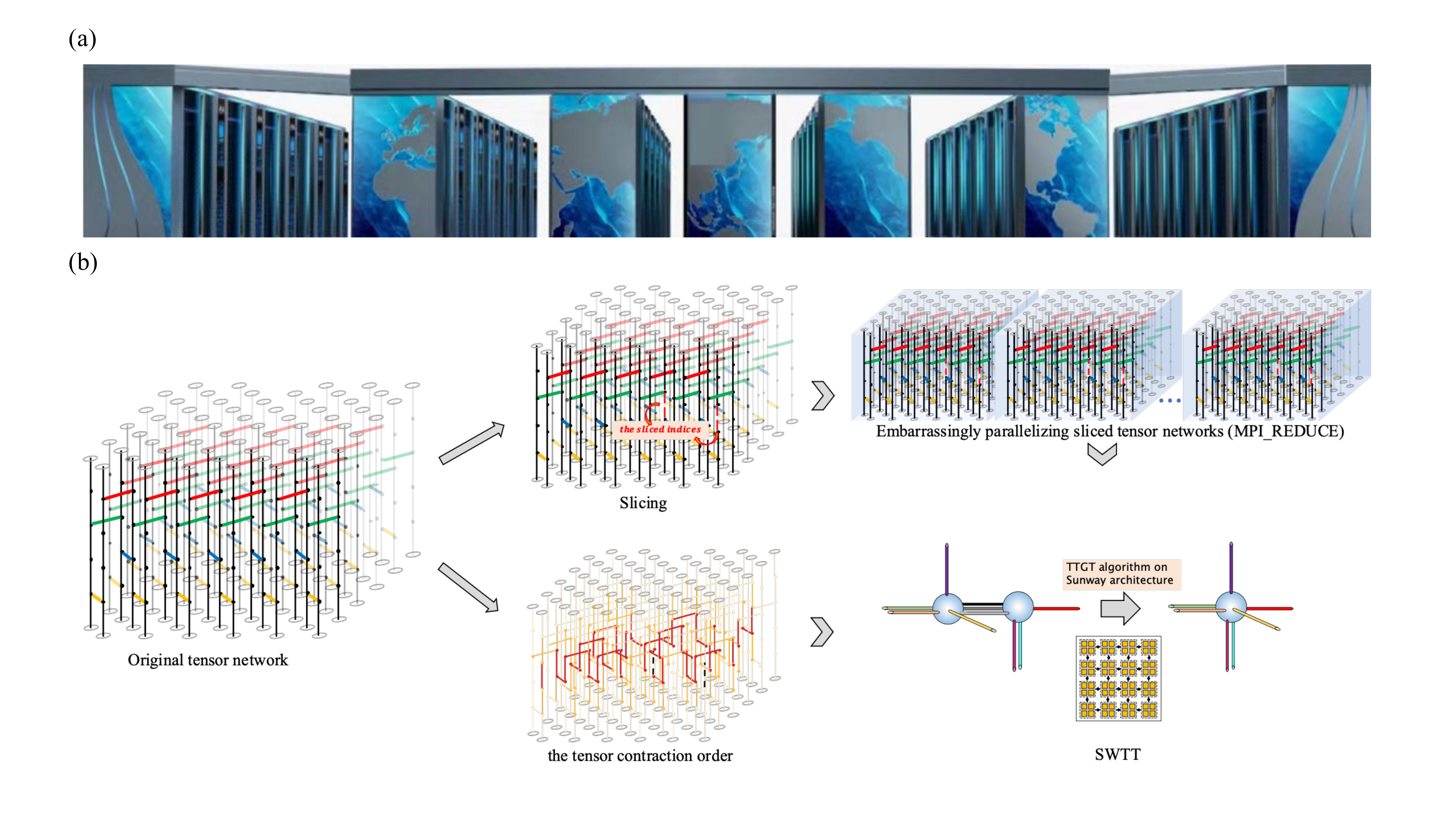}
\caption{ (a) A rendered photo of the new generation Sunway supercomputer. 
(b) \gcc{Demonstration of the two-level parallelization scheme. The tensor network on the left-most side is first sclied into many smaller tensor networks, each of which is distributed onto a single CPU. Then for each smaller tensor network the pair-wise tensor contractions are done with our fused tensor permutation and multiplication algorithm, which is further parallelized on the $384$ cores of a single CPU.}
}
\label{fig:fig1}
\end{figure*}

\section{Efficient implementation of a customized TNC algorithm}

Mathematically, an $n$-qubit quantum state is represented as a rank-$n$ tensor, with $2^n$ entries of complex numbers, while single-qubit and two-qubit quantum gate operations can be represented as rank-$2$ and rank-$4$ tensors respectively. Applying a gate operation amounts to contracting the common tensor indices between the gate operation and the quantum state. This is the so-called Schr$\ddot{\text{o}}$dinger algorithm with a time complexity $O(m2^n)$ and space complexity $O(2^n)$ for a quantum circuit with $m$ gate operations. For Sycamore-$20$, we have $n=53$ and $m \approx 400$ (single-qubit gate are neglected since they can be absorbed into two-qubit gates) and thus the time complexity is $O(10^{18})$. Ideally, this could be accomplished within seconds on an exascale supercomputer! The real limitation is the memory cost to store the quantum state, which is about $68$ petabytes if stored as single-precision complex numbers and is at least one order of magnitude larger than the memory size of state of the art supercomputers. The implementation of the Schr$\oo$dinger algorithm up to now is thus limited within $45$ qubits~\cite{DeIto2007,SmelyanskiyGuzik2016,HanerSteiger2017,PednaultWisnieff2017}.

To simulate the Sycamore quantum processor or beyond, an algorithm to trade time for space has to be used. Here we note the work from IBM researchers that leverages secondary storage to overcome the memory issue, which however is a more of a theoretical proposal instead of actual implementation~\cite{IBM2019}. During the past two years the method of choice to simulate the Sycamore-like quantum processor has gradually converged to a specific type of TNC algorithm~\cite{MarkovShi2008,HuangChen2021,PanZhang2021}, which treats the gate operations, the initial quantum state as well as the target computational basis (which are both made of separable rank-$1$ tensors) as a whole tensor network. Contracting this tensor network results in the exact amplitude for this basis. Such tensor network contraction is in general an NP-hard problem, whose performance greatly relies on a properly devised tensor contraction order. TNC is often used in conjunction with a \textit{tensor index slicing} scheme, which slices a number of tensor indices such that the original tensor network is equivalent to the summation of a bunch of smaller tensor networks. This technique is important in practice since the largest intermediate tensor during the contraction of the original tensor network could easily exceed the available amount of memory for large problems. As a result, the complexity of the original tensor network is that of each sliced tensor network times the total number of slices. Researchers from Alibaba propose an intertwined tensor slicing and subtree reconfiguration scheme to optimize the slicing and the tensor contraction order for the sliced tensor network altogether~\cite{HuangChen2021}, whose strategy is also adopted in latest version of the package cotengra~\cite{GrayKourtis2020}. Currently, the tensor contraction order found for Sycamore-$20$ using different strategies are more or less of the same order of $10^{18}$~\cite{HuangChen2021,PanZhang2021,GrayKourtis2020} and in this work we use cotengra to produce a near-optimal tensor contraction order for later computation.


Given a specific tensor contraction order, the performance of TNC then depends mostly on the performance of pair-wise tensor contraction. The new generation Sunway supercomputer is powered by the homegrown SW26010P CPU, each with a total of $96$ GB memory and is further divided into $6$ core groups, each with $64$ cores. The single-precision and half-precision performances are $14$ TFLOPS and $53$ TFLOPS respectively, and the memory bandwidth is $307$ GB/s. Unfortunately, we find in practice that the resulting TNC produced by cotengra (also similar for other approaches) is dominated by highly skewed tensor contractions, namely contraction between a very high-rank tensor and a very low-rank tensor, for which the floating point arithmetic complexity is essentially of the same order as the memory access complexity (both proportional to the size of the larger tensor). As a result the ultimate performance of the tensor contraction is limited by the memory bandwidth. 
Nevertheless, to make the most utilization of our CPU architecture, we propose a fused tensor permutation and multiplication algorithm to push the efficiency of skewed tensor contraction to its extreme.

Our full-scale implementation starts with a two-level parallelization scheme which demonstrates to be most efficient for Sycamore-$20$: the inner level consists of $6$ core groups (a single CPU) which takes a single sliced tensor network as input and distributes each tensor contraction onto the $384$ cores; the outer level distributes all the slices into different CPUs.
The tensor index slicing is done with a maximally allowed intermediate tensor size of $2^{31}$ for single precision, which results in $2^{22}$ slices, and $2^{30}$ for mixed precision, which results in $2^{23}$ slices. For the highly skewed pair-wise tensor contraction performed inside each CPU between a high-rank tensor $A$ and a low-rank tensor $B$, we first store a copy of $B$ in the local data memory (LDM) of each CPU core ($256$ KB), and then we fetch the slices of $A$ into LDMs of each cores and perform a local matrix multiplication in parallel. The central design principle is that $A$ and $B$ are only loaded into the LDMs once. 
\gcc{The parallelization scheme as well as the major steps of the tensor network contraction algorithm are also shown in Fig.~\ref{fig:fig1}(b) (see Supplementary for more details). }
In practice we find that for typical tensor networks generated for the quantum supremacy circuits, our implementation can reach a $0.63$ TFLOPS single-precision efficiency for each CPU (which surpasses the memory bandwidth by a factor of $2$ due to the existences of a few computational intensive tensor contractions). Compared to the commonly used package jax~\cite{jax2018github}, there is an average speedup of more than $25{\rm x}$ and a maximum speedup of more than $100{\rm x}$. The near optimal software implementation together with the unprecedented parallelization scale over $41,932,800$ cores allow us to tackle those quantum supremacy circuits which are believed to be extremely difficult or impossible to simulate previously.



To this end we note that the TNC algorithm used here is in sharp comparison with the ones used in Ref.~\cite{GuoWu2019,VillalongaMandra2018,VillalongaMandra2019,GuoHuang2021}, where the tensors in the time direction are pre-contracted, resulting in a two-dimensional tensor network with the same geometry as the quantum processor. In the latter case it has been shown that the FLOPS efficiency could easily exceed $60\%$~\cite{LiuChen2021}. However for Sycamore-$20$, each tensor in the resulting two-dimensional tensor network is too large to be stored on a single CPU, making it impractical for quantum supremacy circuits with more than $14$ cycles~\cite{AruteMartinisQuantumSupremacy2019}.

\section{Establishing quantum supremacy baseline}

\begin{figure}
\includegraphics[width=\columnwidth]{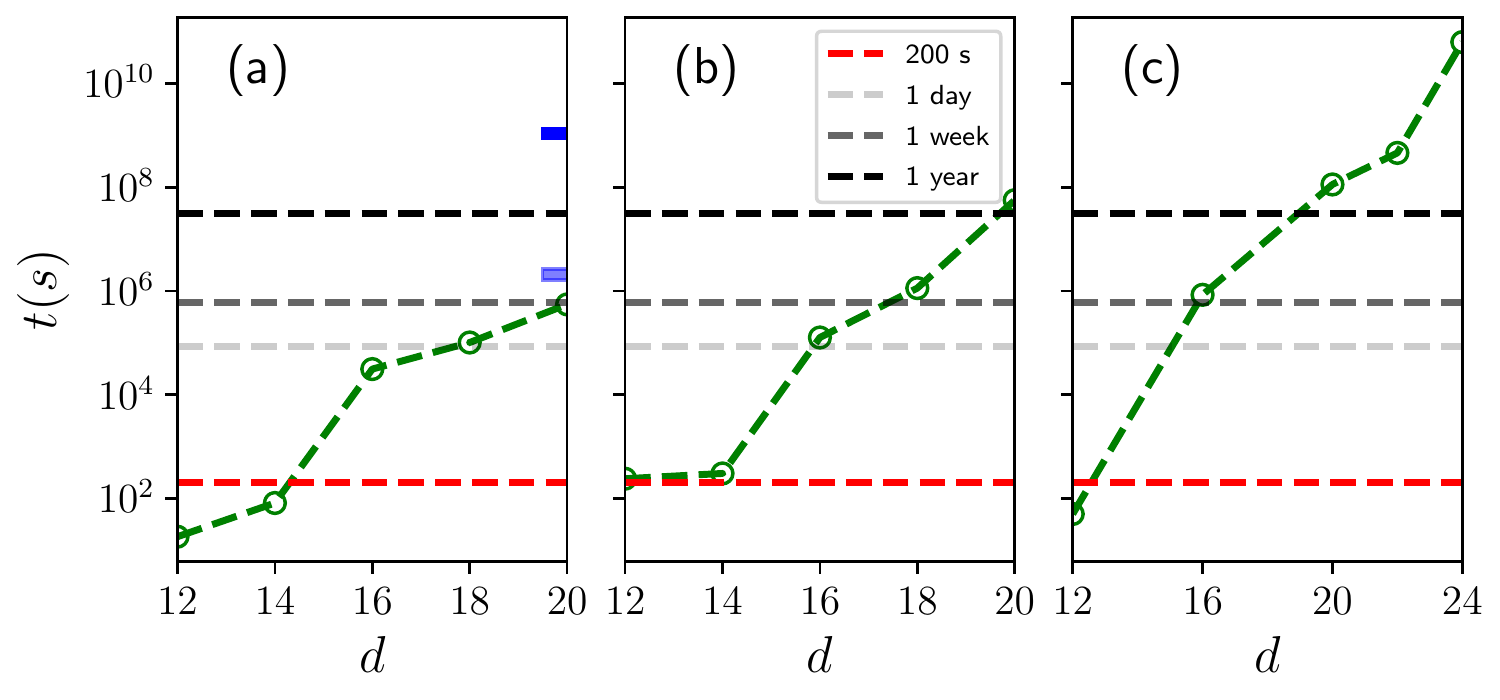}
\caption{ Total runtime (green dashed line with circle) to simulate the same sampling task (producing $1$ million bitstrings with the same XEB fidelity as the quantum processor) for quantum supremacy circuits on (a) Sycamore (b) Zuchongzhi $2.0$ and (c) Zuchongzhi $2.1$, based on extrapolation from the runtime to generate a single perfect sample. The x-axis $d$ denotes the number of cycles. For Sycamore and Zuchongzhi $2.0$, we direct generate one perfect sample. While for Zuchongzhi $2.1$, we estimate the time to generate a perfect sample from the runtime to compute a single slice using a single CPU. The blue rectangle marks the total runtime for Ref.~\cite{PanZhang2021} on $60$ GPUs while the light blue rectangle marks the estimated total runtime for Alibaba~\cite{HuangChen2021} on Summit.
}
\label{fig:fig2}
\end{figure}

The classical runtime of the quantum supremacy circuits are estimated by evaluating the runtime for generating one perfect sample (except in the case of Zuchongzhi $2.1$ where sampling a single bitstring classically already becomes demanding). This is done by computing a batch of $64$ amplitudes in a single run by leaving $6$ qubits open and then perform frugal sampling, which could guarantee to produce one sample with probability close to $1$~\cite{VillalongaMandra2018,HuangChen2021}. Here we note that for TNC algorithm there is a very economic way to compute a correlated bunch of amplitudes in a single run by reusing the contraction outcome of a major portion of the tensor network, and iterate over the rest small portion. In this way, the overhead of computing a small bunch of correlated amplitudes is negligible compared to computing a single amplitude. 
Moreover, it has been shown that to match the sampling outcomes from a noisy quantum computer with an XEB fidelity $f$, the classical simulation cost can be simply reduced by a factor of $f$~\cite{MarkovBoixo2018}. As a result, the classical complexity of generating $1$ million samples with $0.2\%$ XEB fidelity is equivalent to generating $2000$ perfect samples.

The estimated runtimes for the Sycamore supremacy circuits are shown in Fig.~\ref{fig:fig2}(a). In particular, for Sycamore-$20$ our runtime for generating a perfect sample is $276$ seconds, and thus generating $2000$ prefect samples (or equivalently $1$ million samples with $0.2\%$ fidelity) would only requires $6.4$ days, which could be straightforwardly simulated with the new generation Sunway supercomputer. This is the fastest record till now, which clearly collapses the claim of quantum supremacy for the Sycamore processor. Moreover, the runtimes for Sycamore-$12$ and Sycamore-$14$ are only $18$ and $82$ seconds respectively (their estimated values in Ref.~\cite{AruteMartinisQuantumSupremacy2019} are $2$ hours and $2$ weeks respectively), which are even faster than Sycamore ($200$ seconds). 

In Fig.~\ref{fig:fig2}(b,c) we also show our estimated runtimes for Zuchongzhi $2.0$ and Zuchongzhi $2.1$ respectively. We can see that Zuchongzhi $2.0$-$20$ and Zuchongzhi $2.1$ with beyond $20$ cycles require runtimes of more than $1$ year, which are currently beyond our reach. The most difficult candidate is Zuchongzhi $2.1$-$24$, for which generating a single perfect sample would require a classical runtime of around $5$ years for us.

\begin{figure*}
\includegraphics[width=2\columnwidth]{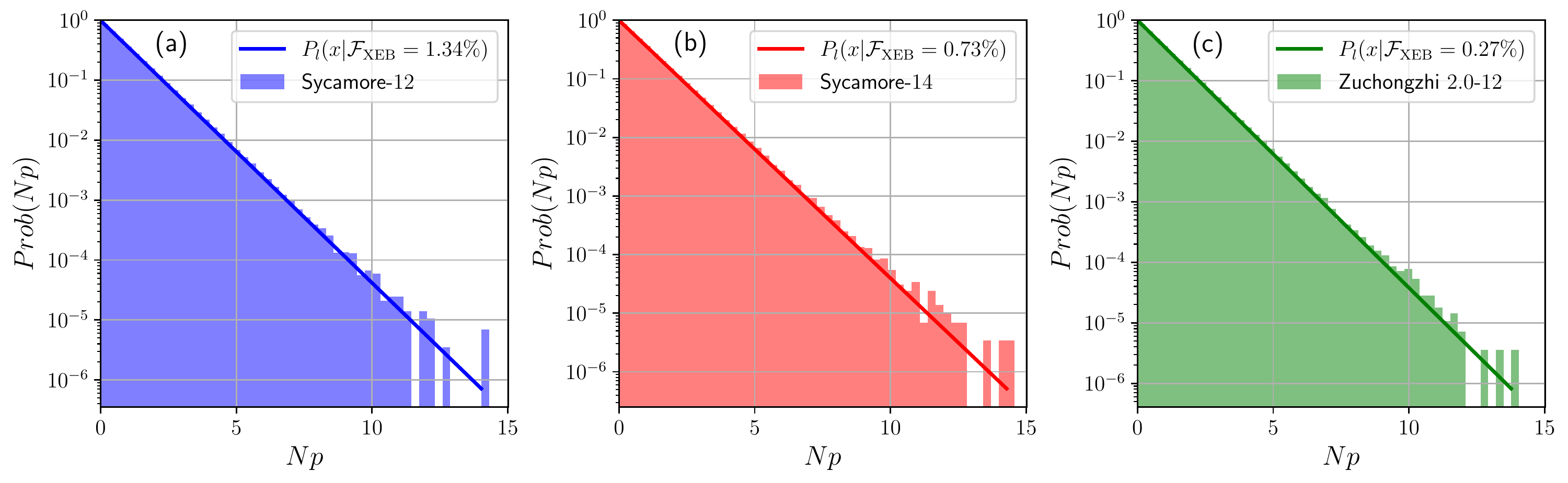}
\caption{ Histograms for the distributions of the amplitudes corresponding to the experimentally generate bitstrings for (a) Sycamore-$12$, (b) Sycamore-$14$ and (c) Zuchongzhi $2.0$-$12$. The solid lines denote the corresponding theoretically predictions with the same XEB fidelities as from Eq.(\ref{eq:fxeb}).
}
\label{fig:fig3}
\end{figure*}

\section{Verifying quantum supremacy circuits}

The derivation of the XEB fidelities for the quantum supremacy circuits is slightly subtle since to obtain the XEB fidelities one needs to compute amplitudes of the experimentally generated bitstrings with classical computers, which is considered not possible for those circuits. 
In fact, the estimated runtimes for computing $1$ million exact amplitudes for Sycamore-$12$ and Sycamore-$14$ are $6$ days and $4.26$ years respectively~\cite{AruteMartinisQuantumSupremacy2019}. The XEB fidelities of those quantum supremacy circuits are then estimated based on two different approaches: 1) extrapolation based on component level fidelities which assumes a good suppression of the cross talks between gate operations and 2) extrapolation based on simplified variants for which a small portion of the gate operations is removed. The elided circuits are used for the latter purpose. Taking the elided version of Sycamore-$20$ for example, the number of removed gate operations is less than $30$, which is small compared to the total number of gate operations which is more than $1000$. However, the amount of entanglement in the underlying quantum states is reduced by more than half after removing those gates (thus the classical simulation complexity decreases exponentially). The equivalence between the XEB fidelities of the elided circuits and the corresponding supremacy circuits thus may not be easily justified.

Nevertheless, the XEB fidelity is a vital ingredient to characterize those near-term quantum processors. In the first place, the quantitive comparison between quantum and classical computers is only well-defined once the XEB fidelity of the quantum sampling process is precisely known. With our highly efficient random quantum circuit simulator on the new generation Sunway supercomputer, it is possible to directly compute a large number of exact amplitudes for quantum supremacy circuits with less than $14$ cycles, with which we can compute the exact XEB fidelities and verify them against the estimated ones. Concretely, we compute a million exact amplitudes for Sycamore-$12$, Sycamore-$14$ and  Zuchongzhi $2.0$-$12$, from which we obtain their XEB fidelities as $(1.34\pm 0.1)\%$, $(0.73\pm 0.1)\%$ and $(0.27\pm 0.1)\%$ respectively, in comparison with the values of $1.4\%$, $0.9\%$ and $0.37\%$ as estimated from their elided versions. We can see that the exact XEB fidelities for those quantum supremacy circuits are slightly lower than their estimated values, but still within errorbar, which may be due to the reason that their estimated values are computed using a lot more bitstrings than $1$ million.

We plot in Fig.~\ref{fig:fig3} the histograms of those amplitudes and compare them to the theoretical probability density function (PDF) for the rescaled bitstring probability $Np$ ($N=2^n$ and $p$ is the probability) under the same XEB fidelity $\fxeb$~\cite{AruteMartinisQuantumSupremacy2019}, namely 
\begin{align}\label{eq:fxeb}
P_l(x|\fxeb) = \left(\fxeb x + (1-\fxeb) \right) e^{-x},
\end{align}
with $x=Np$. We can see that they agree well with each other, therefore the bitstrings generated with those quantum supremacy circuits indeed obeys Porter-Thomas distribution with corresponding XEB fidelities. This result provides a strong consistency check for those quantum supremacy experiments which could also be one of the major application scenarios for highly efficient classical simulators of quantum circuits in developing next-generation quantum processors.

\section{Discussion}
The motivation for developing efficient classical simulators for quantum circuits is ultimately driven by the lack of logical qubits and the existence of cross talks in current quantum processors. The Sycamore and Zuchongzhi series of quantum processors reinforce the need to explore the classical computing capacity to its extreme, either for performance benchmarking or fidelity verification. 

Although it is believed that the quantum processors will become exponentially more difficult to be simulated classically, our results together with other works during the last two years after Sycamore, seem to indicate that classical computational capacity for simulating random quantum circuits is growing in pace with the quantum counterparts. This is partially due to that the new Zuchongzhi series quantum processors grow more in terms of the number of qubits but less of the gate operation fidelities~\cite{ZlokapaLidar2020}, and that the classical simulation algorithm, especially TNC based algorithms and classical accelerators are also progressing rapidly.

Our results herald a renewed starting of the quantum-classical competition in terms of simulating quantum supremacy experiments in exascale systems. There is still plenty of space for optimization on the classical side. In the first place, the memory bandwidth and the half-precision performance are apparently bottlenecks for our simulations, which are only $20 \%$ and $17\%$ of that of NVIDIA A100 GPU. Since our current approach is essentially bounded by the memory bandwidth, we expect an immediate $5{\rm x}$ performance increase using other CPU architectures with higher memory bandwidths. A more intelligently devised tensor contraction order could certainly help, especially if the balance between computation and memory could be better taken into account. Additionally, it is recently shown that to compute a large number of uncorrelated amplitudes it is still possible to reuse a large number of intermediate tensors and easily gain a $20{\rm x}$ performance increase~\cite{KalachevYung2021}. We thus believe that in near term the time to classically simulate Sycamore-$20$ would be reduced by $2$ orders of magnitude (about $1.5$ hours). 

\begin{acknowledgments}
This work is partially supported by National Key R$\&$D Program of China (2017YFA0604500), and National Natural Science Foundation of China (U1839206).
C. G acknowledges support from National Natural Science Foundation of China under Grants No. 11805279, No. 61833010, No. 12074117 and No. 12061131011.
H.-L.H. acknowledges support from the Youth Talent Lifting Project (Grant No. 2020-JCJQ- QT-030), National Natural Science Foundation of China (Grant No. 11905294), China Postdoctoral Science Foundation, and the Open Research Fund from State Key Laboratory of High Performance Computing of China (Grant No. 201901-01).
\end{acknowledgments}

\bibliographystyle{apsrev4-1}
\bibliography{refs}

\end{document}


\title{Supplementary Information: Redefining the Quantum Supremacy Baseline With a New Generation Sunway Supercomputer}

\date{\today}

\maketitle




\section{Random quantum circuits}

The random quantum circuits (RQCs) on Sycamore and Zuchongzhi series of quantum processors share a similar structure which consists of interlacing layers of single-qubit gates and two-qubit gates, where each layer of two-qubit gates is counted as one cycle. The single-qubit gates are randomly and independently selected from the set $\{\sqrt{X}, \sqrt{Y}, \sqrt{W}\}$ defined as
\begin{align}
\sqrt{X} &= \frac{\sqrt{2}}{2} \left[ \begin{array}{cc} 1 & -i \\ -i & 1 \end{array} \right]; \\ 
\sqrt{Y} &= \frac{\sqrt{2}}{2} \left[ \begin{array}{cc} 1 & -1 \\ 1 & 1 \end{array} \right]; \\ 
\sqrt{W} &= \frac{\sqrt{2}}{2} \left[ \begin{array}{cc} 1 & -\frac{\sqrt{2}}{2}(1+\im) \\ \frac{\sqrt{2}}{2}(1-\im) & 1 \end{array} \right].
\end{align}
The two-qubit gate is modeled by the $5$-parameter fSim gate 
\begin{align}
\left[\begin{array}{cccc} 1 & 0 & 0 & 0 \\ 0 & e^{\im(\Delta_+ + \Delta_-)}\cos(\theta) & -\im e^{\im(\Delta_+ - \Delta_{-,{\rm off}})}\sin(\theta) & 0 \\ 0 & -\im e^{\im(\Delta_+ + \Delta_{-,{\rm off}})}\sin(\theta) & e^{\im(\Delta_+ + \Delta_-)}\cos(\theta) & 0 \\ 0 & 0 & 0 & e^{\im(2\Delta_+ - \phi)} \end{array} \right]. \nonumber
\end{align}
The values of those parameters are obtained through certain calibration procedures.

\begin{figure}
\includegraphics[width=\columnwidth]{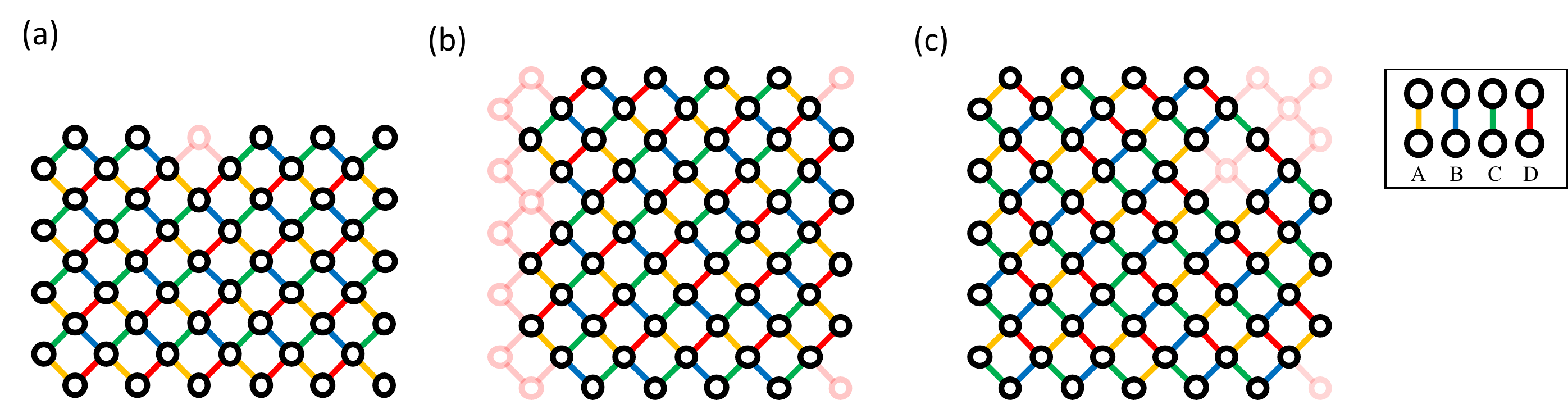}
\caption{Patterns of two-qubit gate layers for (a) Sycamore, (b) Zuchongzhi $2.0$ and (c) Zuchongzhi $2.1$.
}
\label{fig:figS1}
\end{figure}

In all the three quantum supremacy experiments, $4$ different patterns of two-qubit layers are used, which are denoted as $A,B,C,D$. Those $4$ patterns are shown in Fig.~\ref{fig:figS1}, where each empty circle denotes one qubit and  each line between two circles denotes a two-qubit gate operation. The shaded qubits and lines mean that those qubits as well as gate operations are not used in experiments, possibly due to high error rates. As a result each pair of nearest-neighbour qubits is operated on by a two-qubit gate once and only once in each $4$ cycles containing all $A,B,C,D$. A RQC of a fixed number of cycles is then formed by repeating these $4$ patterns, with randomly generated single-qubit gate layers in between. For example a RQC of $8$ cycles is chosen as $ABCDCDAB$, and a RQC of $12$ cycles is $ABCDCDABABCD$. RQCs with cycles which are not divided by $4$ are implemented slightly differently for Sycamore and for Zuchongzhi series and we will not discuss about those details here. The overall design principle is to make the classical simulation cost as high as possible given the same number of qubits and the same amount of quantum gate operations. 

In practice, the number of bitstrings generated experimentally should be larger than a minimal number to ensure that the uniform sampling with $0$ XEB fidelity is rejected with certain level of confidence. This minimal number is thus related to the overall fidelity of the underlying RQC. As a result the number of samples required by the Zuchongzhi series experiments are larger than that required in the Sycamore experiments due to the lower fidelities, and that the number of samples should also be larger for RQCs with more cycles. However, here we simply choose a fixed number, that is, $1$ million, for all those experiments for a simple benchmarking since the complexity of sampling from the quantum processor and from our classical simulator (based on TNC algorithm) both grow linearly with the number of required samples (bitstrings). 


\section{Component-level calibration}
Ideally, one could calibrate each single-qubit gate and two-qubit gate and then the fidelity of a full RQC can be estimated as the product the fidelities of all the component gates. However due to the cross talks among different gate operations the fidelities of the RQCs estimated in this way would be lower than the actually measured values. Ref.~\cite{AruteMartinisQuantumSupremacy2019} proposes a component-level calibration scheme to characterize the fidelities of the two-qubit gate operations, which is demonstrated to have better generalizability when used to predict fidelities of the full RQCs. This scheme is also adopted in Zuchongzhi $2.0$~\cite{WuPan2021} and will be briefly reviewed in the following.

To characterize the two-qubit gate XEB fidelities for the two-qubit gate operations belonging to pattern $A$, for example, one use a circuit which only contains repeating layers of the $A$ patterns (with single-qubit gate layers in between), and then measure all the XEB fidelities of these pairs of two qubits altogether. After subtracting the contributions of the single-qubit gate operations, whose XEB fidelities are characterized before hand, one obtains the XEB fidelities of all those two-qubit gate operations belonging to pattern $A$ in a parallelized fashion. It turns out that the XEB fidelities obtained in this way for the two-qubit gate operations are slightly lower than those obtained from individual calibrations, but they can better predict the XEB fidelities of the full RQCs since the effects of cross talks have been absorbed into those fidelities to a certain level. In Ref.~\cite{ZhuPan2021}, however, an even more elaborated calibration scheme is used and we will not go into the details of it here. The main message here is that for current noisy quantum processors, it is difficult to completely get rid of the effects of cross talks. As a result, the full-circuit level benchmarking of the fidelities will be important, either as a brute-force calibration scheme itself, or as a consistency check for the more elaborated subsystem calibration schemes in the future.

\section{Tensor network contraction algorithm}

\begin{figure}
\includegraphics[width=0.6\columnwidth]{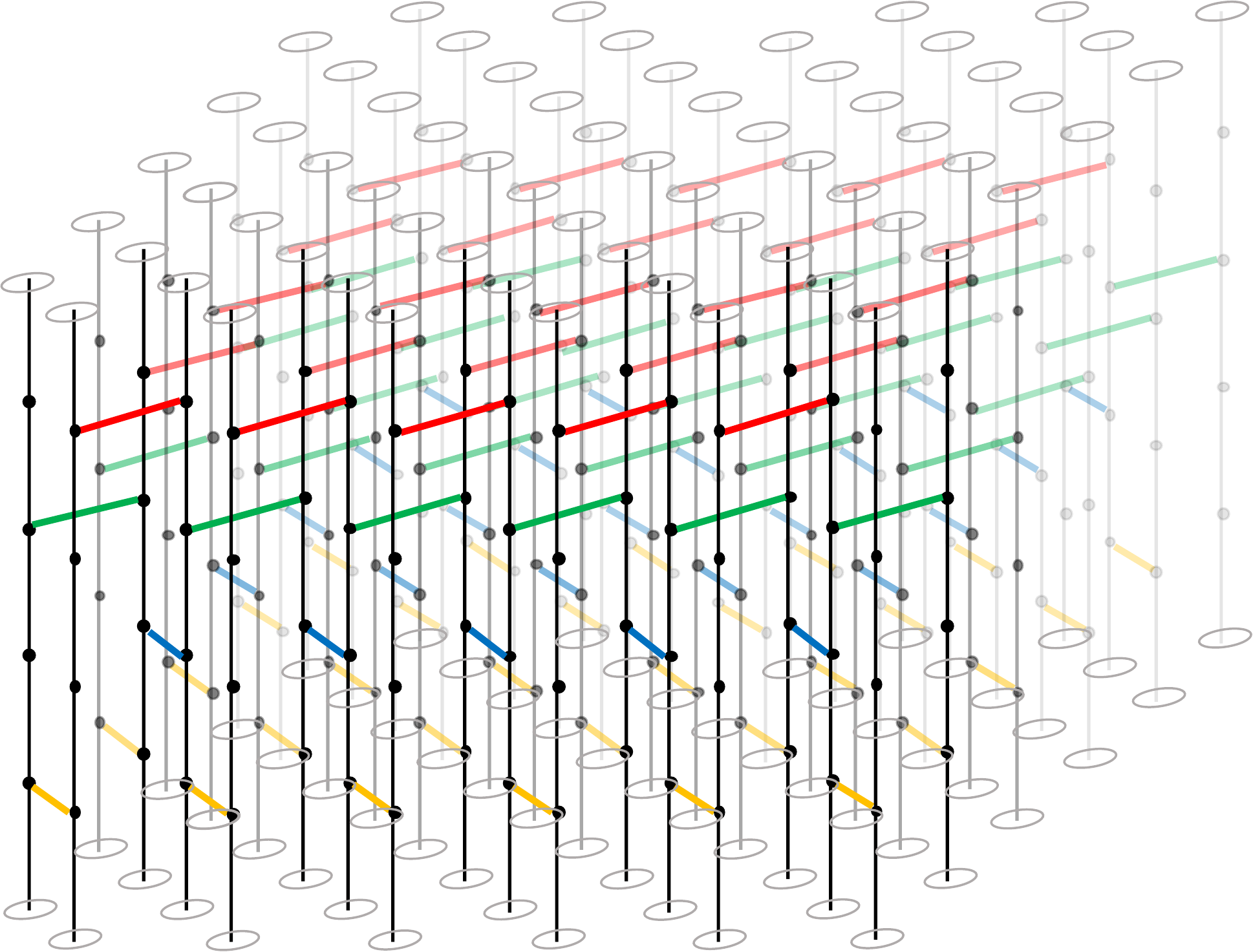}
\caption{Map of the task of computing a single amplitude into contracting a tensor network. The vertical direction denotes the time direction (from bottom up). The black lines separated by black balls denote tensor legs, each with dimension $2$. The empty circles in the bottom and top layers represent the rank-$1$ tensors resulting from the initial state and final computational basis respectively. The black balls without any colored edges between them denote single-qubit gate operations. The colored edges together with two connected black balls denote two-qubit gate operations, where different colors denote two-qubit gate operations belonging to different patterns from $A, B, C, D$. 
}
\label{fig:figS2}
\end{figure}

Computing a single amplitude from the output of a quantum circuit can be naturally mapped to the problem of contracting a tensor network, as demonstrated in Fig.~\ref{fig:figS2}. The tensor network can also be viewed as a graph, which is assumed to be connected, since otherwise it means that the original RQC can be partitioned into two independent smaller quantum circuits, which could then be simulated separately with much lower classical costs. Since rank-$1$ (resulting from the initial state and the final computation basis) and rank-$2$ tensors (single-qubit gates) can always be pre-absorbed into higher-rank tensors without increasing the sizes of tensors (two tensors with ranks larger than $2$ will also be pre-contracted if the rank of resulting tensor is not larger than one of these two tensors), the tensor network will thus always be simplified to an equivalent one which only contains rank-$3$ tensors or above. As an example, for Sycamore-$20$ which contains a total of more than $430$ two-qubit gate operations, the resulting tensor network contains less than $400$ nodes, with each node a rank-$4$ tensor.

The complexity of contracting a tensor network is heavily dependent on the underlying tensor contract order, which is a list of two-tuples specifying the sequence of pair-wise tensor contractions. However, finding the optimal tensor contraction order for an arbitrary graph is an NP-hard problem and is almost impossible for our case with hundreds of nodes. Important progresses have been made in the last two years to find near optimal tensor contraction orders with heuristic algorithms~\cite{GrayKourtis2020,HuangChen2021,PanZhang2021}. In this work we combine a set of heuristic algorithms to produce a near optimal tensor contraction order for later processing. Concretely, we use the hypergraph partitioning algorithm from the package kahypar to decompose the initial tensor network into several connected subtrees, each of which is itself a smaller tensor network~\cite{kahypar}, and then we iteratively slice those subtrees and optimize each subtree (this is cheap and can be done by brute force since the number of nodes in the subtree is usually quite small) to reduce the the tensor network contraction complexity under a given memory bound for the largest intermediate tensor. We list the floating point arithmetic complexity corresponding to the tensor contraction orders found for several quantum supremacy circuits in TABLE.~\ref{tab:tab1}.

\begin{table}[!htb]
\centering
\caption{Floating point arithmetic complexities corresponding to the best tensor contraction orders found for several quantum supremacy circuits. Here the largest allow size for the intermediate tensors is $2^{32}$.
}
\label{tab:tab1}
\begin{tabular}{|c|c|c|}
\hline
circuit & number of slices & complexity  \\
\hline
Sycamore-$20$ & $2^{21}$ & $6.92 \times 10^{18}$ \\
Zuchongzhi $2.0$-$20$ & $2^{29}$ &  $1.29 \times 10^{21}$ \\
Zuchongzhi $2.1$-$20$ & $2^{29}$ &  $2.55 \times 10^{21}$  \\
Zuchongzhi $2.1$-$24$ & $2^{40}$ &  $2.08 \times 10^{24}$  \\
\hline
\end{tabular}
\end{table}

In practice, the tensor contraction order with a lower floating point arithmetic complexity may not be faster due to the memory intensive nature of the algorithm. To circumvent this problem, we generate $100$ tensor contraction orders and run a single slice on a single CPU for each of them, and then choose the fastest one.

\section{CPU architecture and the tensor contraction algorithms}

\begin{figure}
\includegraphics[width=0.8\columnwidth]{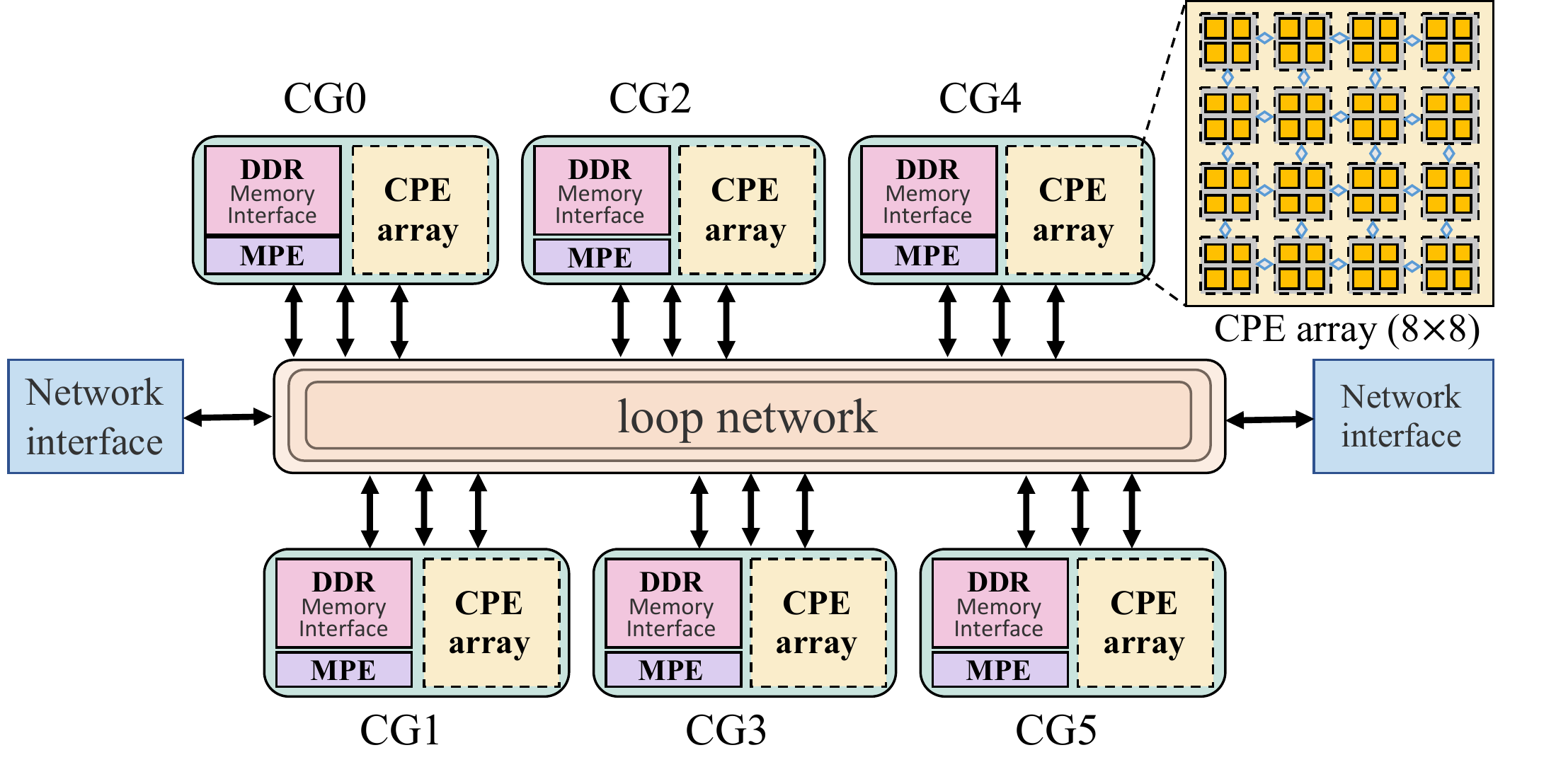}
\caption{Architecture of the SW26010P CPU.
}
\label{fig:figS3}
\end{figure}

The SW26010P CPU contains $6$ core groups (CGs). Each CG contains $8\times 8$ cores and a total of $16$ GB of DDR4 memory with a bandwidth of $51.2$ GB/s. Each core has an LDM of $256$ KB. Data from the DDR4 memory can be loaded into the LDM of each core through the direct memory access (DMA), and different cores inside the same CG can interchange data through remote memory access (RMA). In total, a single SW26010P CPU has $384$ cores and $96$ GB memory, a $14$ TFLOPS single-precision performance and a $53$ TFLOPS half-precision performance, as well as a $307$ GB/s memory bandwidth. The architecture of a single SW26010P CPU is shown in Fig.~\ref{fig:figS3}. 

In our case the tensor network contraction is dominated by highly skewed pair-wise tensor contractions for which the memory access complexity is roughly equal to the floating point arithmetic complexity. However the memory bandwidth is only around $2.2 \%$ percent of the single-precision performance, as a result the performance of TNC is essentially bounded by the memory bandwidth and we have made an extreme effort to devise a fused tensor permutation and multiplication algorithm to make the best use of the limited memory bandwidth. 

\begin{figure}
\includegraphics[width=0.6\columnwidth]{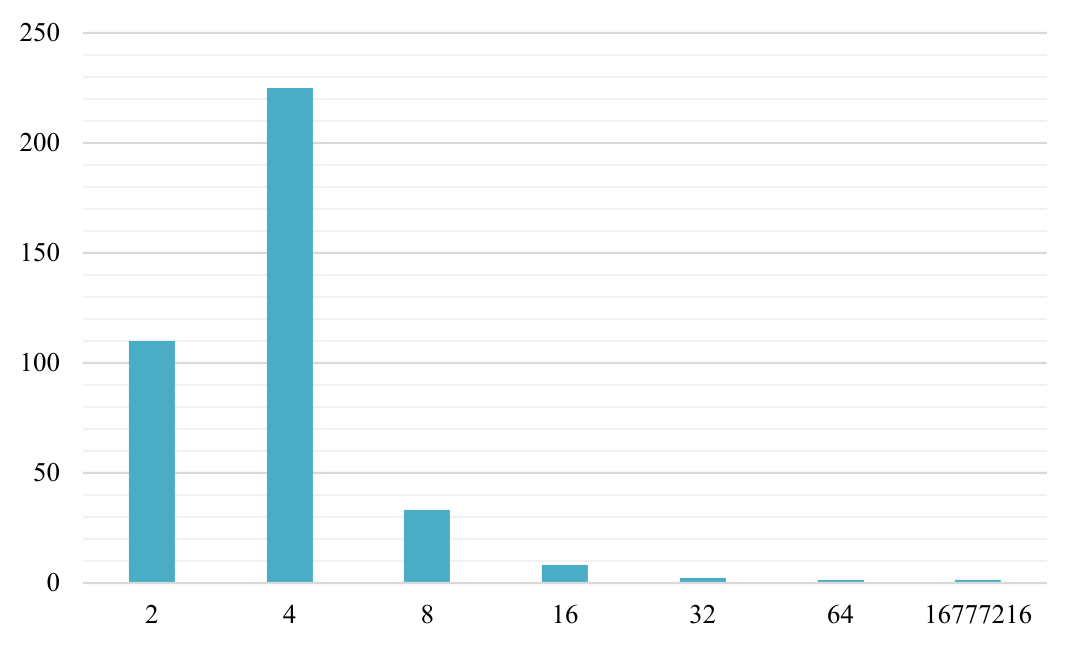}
\caption{Histogram of the contracted size $k$ (x-axis) during the pair-wise contraction of a typical tensor network for Sycamore-$20$.
}
\label{fig:figS4}
\end{figure}

In the next we take the case of a pair-wise tensor contraction between a rank-$6$ tensor $A[a,b,c,d,e,f]$ with a rank-$5$ tensor $B[a, h,i,e,j]$ as an example, with the two common tensor indices $a, e$ to be contracted. The result is a rank-$7$ tensor denoted as $C[b,c,d,f,h,i,j]$. A commonly used (and one of the most efficient) algorithm for tensor contraction is to permute the tensor indices of $A$ and $B$ and then perform matrix multiplication as
\begin{align}\label{eq:matmul}
C[(b,c,d, f), (h,i,j)] = \sum_{(a, e)} A[(b,c,d,f), (a, e)] \times B[(a, e), (h, i,j)],
\end{align}
where $(x, y, z)$ means to fuse the three individual tensor indices $x, y, z$ into a single tensor index. For this algorithm, the permutation requires to load $A$ and $B$ once, which means to load the elements of them from the DDR4 memory into LDMs of each core, perform the permutation locally and then put the results back into DDR4 memory. After that, the following matrix multiplication would require to load the the permuted tensors once again. As a result, each element of $A$ and $B$ are loaded twice. This is ok if the resulting matrix multiplication is more or less balanced, that is, the first matrix has a shape $p\times k$ and the second matrix has a shape $k\times q$, with $p\approx k\approx q$, since in this case the matrix multiplication complexity is $O(k^3)$ while the memory access complexity is $O(k^2)$. However, in our case, as indicated in Eq.(\ref{eq:matmul}), we have $p\gg k \approx q$ and thus both the matrix multiplication complexity and the memory access complexity is $O(p)$, taking into consideration that the memory bandwidth is much smaller than the floating point arithmetic performance, the complexity of the tensor multiplication would be dominated by the memory access. In Fig.~\ref{fig:figS4}, we plot the distribution of $k$ appeared in the pair-wise tensor contractions corresponding to a particular tensor contraction order found for Sycamore-$20$.

\begin{figure}
\includegraphics[width=0.8\columnwidth]{FigS7.pdf}
\caption{(a) Fuse tensor permutation and multiplication algorithm. The indices of $C$ is permuted compared to the natural order in Eq.(\ref{eq:matmul}). (b) Ring RMA scheme when the size of $B$ does not fit into the LDM of each core. We have assumed a column-major storage for all the tensors.
}
\label{fig:figS7}
\end{figure}

Our fused tensor permutation and multiplication algorithm aims to reduce the memory access of $A$ and $B$ to only once. The main idea is as follows. First, since $B$ is assumed to be very small, it is directed loaded into the LDMs of each cores through DMA. Then for a particular configuration of $(b, c, d, f)$, say $(0,1,0,1)$, we load $A[:, 0,1,0,:,1]$ into the LDM of a particular core through DMA, which is a vector of size $4$ (or a matrix of size $2\times 2$). Then one can perform the tensor multiplication kernel locally which is simply a $8\times 4$ matrix (corresponding to $B$) times a vector of size $4$, and the result is stored into $C[0,1,0,1,:,:,:]$. Different configurations of $(b, c,d, f)$ can be loaded into different cores on the same CPU simultaneously with no data communication at all (the same for the local matrix multiplication kernel), thus can be done in parallel. In practice, it is also very helpful to aggregate the data transfer and the computation, For example, we can combine $(a, e)$ with $(b, f)$, namely each time we will fetch a batch of $2^4$ elements from $A$ into an LDM such that the $4$ numbers corresponding to the last two indices $(e, f)$ are fetched contiguously (In our actual implementation on Sunway the smallest batch size is chosen to be $2^{13}$ instead of $2^4$), and then the local matrix-vector multiplication becomes a matrix-matrix multiplication (a $8\times 4$ matrix times a $4\times 4$ matrix). This fused tensor permutation and multiplication algorithm is shown in Fig.~\ref{fig:figS7}(a). In case $B$ does not fit into LDM (which only happens for the Zuchongzhi series), $B$ is further sliced and each LDM stores a single slice, then the different cores interchange their slices using remote memory access (RMA), which is shown in Fig.~\ref{fig:figS7}(b).

\begin{figure}
\includegraphics[width=0.6\columnwidth]{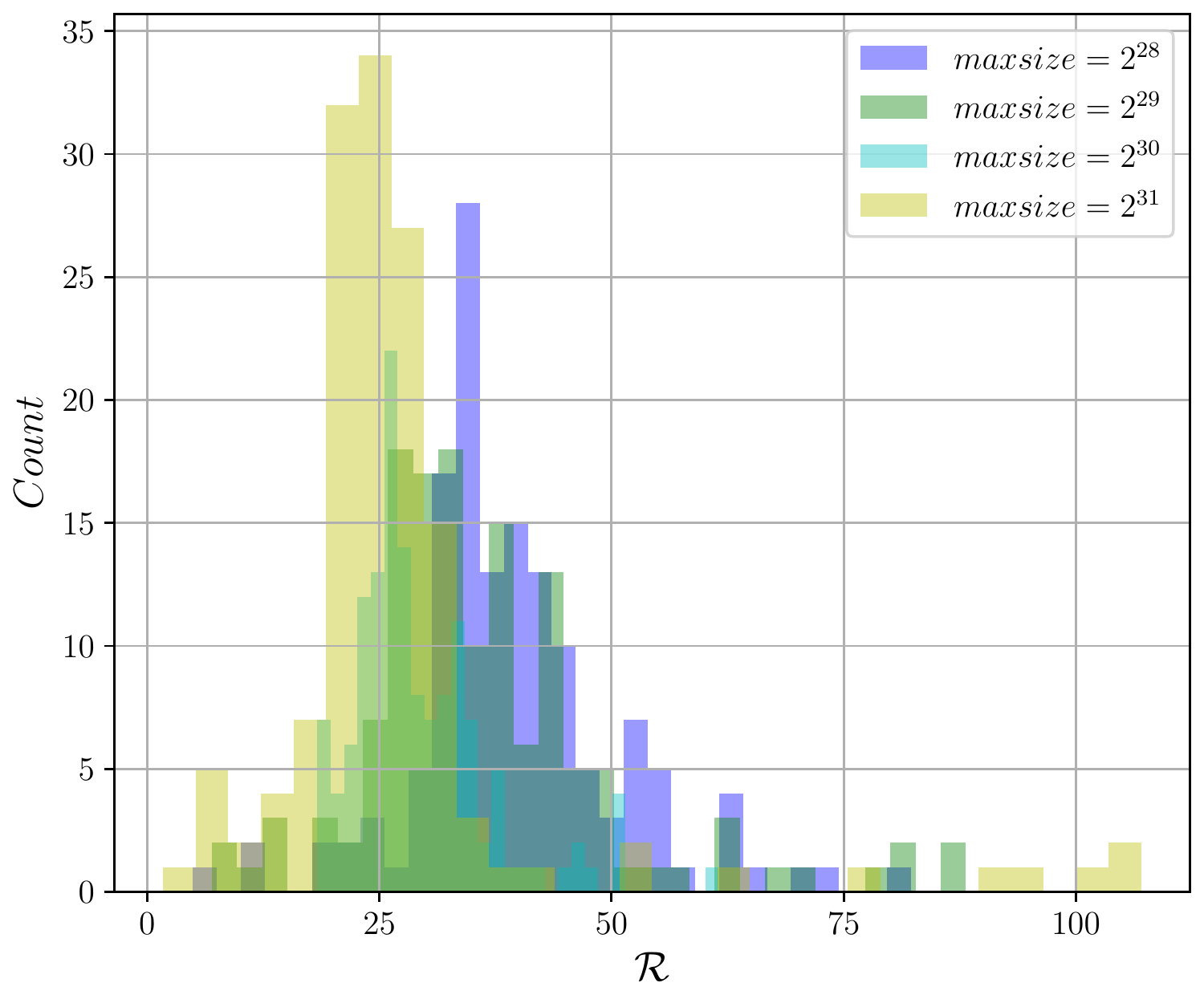}
\caption{Histogram of speedup of our optimized tensor contraction algorithm compared to the package jax~\cite{jax2018github}. Up to $10$ best tensor contraction orders for the quantum supremacy circuits of Sycamore and Zuchongzhi series quantum processors are used, and different thresholds from $2^{28}$ to $2^{32}$ for the size of largest allowed intermediate tensors are considered.  $\mathcal{R}$ is the runtime of jax over our customized tensor contraction algorithm.
}
\label{fig:figS5}
\end{figure}

The speedup of the tensor contraction algorithm compared to the commonly used library jax is shown in Fig.~\ref{fig:figS5}, where we select up to $10$ best tensor contraction orders found for the quantum supremacy circuits of all the three quantum processors and then compare the runtime for contracting a single slice using jax library or our home-made tensor contraction algorithm. We find an average speedup of more than $25{\rm x}$ and a maximum speedup of more than $100{\rm x}$ for certain tensor contraction orders.

\section{Error and scaling analysis}
As a consistency check for our large-scale simulation results, we compute the $5$ amplitudes for the Sycamore processor of $20$ cycles with $5$ particular computational basis as listed in Ref.~\cite{PanZhang2021}, which is shown in TABLE.~\ref{tab:tab2}. The relative error between Ref.~\cite{PanZhang2021} and ours is within $2\%$. Here we we compute those $5$ amplitudes one by one, in comparison with Ref.~\cite{PanZhang2021} which compute them in a single batch.

\begin{table}[!htb]
\centering
\caption{Comparison of $5$ amplitudes corresponding to $5$ bitstrings for Sycamore-$20$ computed by us to the results from Ref.~\cite{PanZhang2021}.
}
\label{tab:tab2}
\begin{tabular}{|c|c|c|c|}
\hline
bitstring & amplitudes (Ref.~\cite{PanZhang2021}) & amplitudes (ours) & relative error \\
\hline
00000000000000000000000100000110100000010000100000000 &   $-2.97\times 10^{-8} + 2.06\times 10^{-8}\im$     &  $-2.97\times 10^{-8} + 2.07\times 10^{-8}\im$ & $0.05\%$ \\
00000000000000000000000000000000000000000100001000000 &   $1.5\times 10^{-8} + 3.85\times 10^{-9}\im $     &  $1.5\times 10^{-8} + 3.85\times 10^{-9}\im $ & $0.07\%$ \\
00000000000111000001111100001111100001111100110000000 &   $-3.17\times 10^{-9} - 5.45 \times 10^{-9}\im $     &  $-3.17\times 10^{-9} - 5.43 \times 10^{-9}\im $ & $0.26\%$  \\
00000000000111000001111100001111100001111000101000000 &   $-1.89\times 10^{-10} + 3.13 \times 10^{-9}\im $     &  $-1.78\times 10^{-10} + 3.12 \times 10^{-9}\im $ & $0.52\%$  \\
00000000000000000000000000000100000000011100010000000 &   $8.07\times 10^{-10} + 4.35 \times 10^{-10}\im $     &  $8.14\times 10^{-10} + 4.2 \times 10^{-10}\im $ & $1.77\%$  \\
\hline
\end{tabular}
\end{table}

\begin{figure}
\includegraphics[width=0.8\columnwidth]{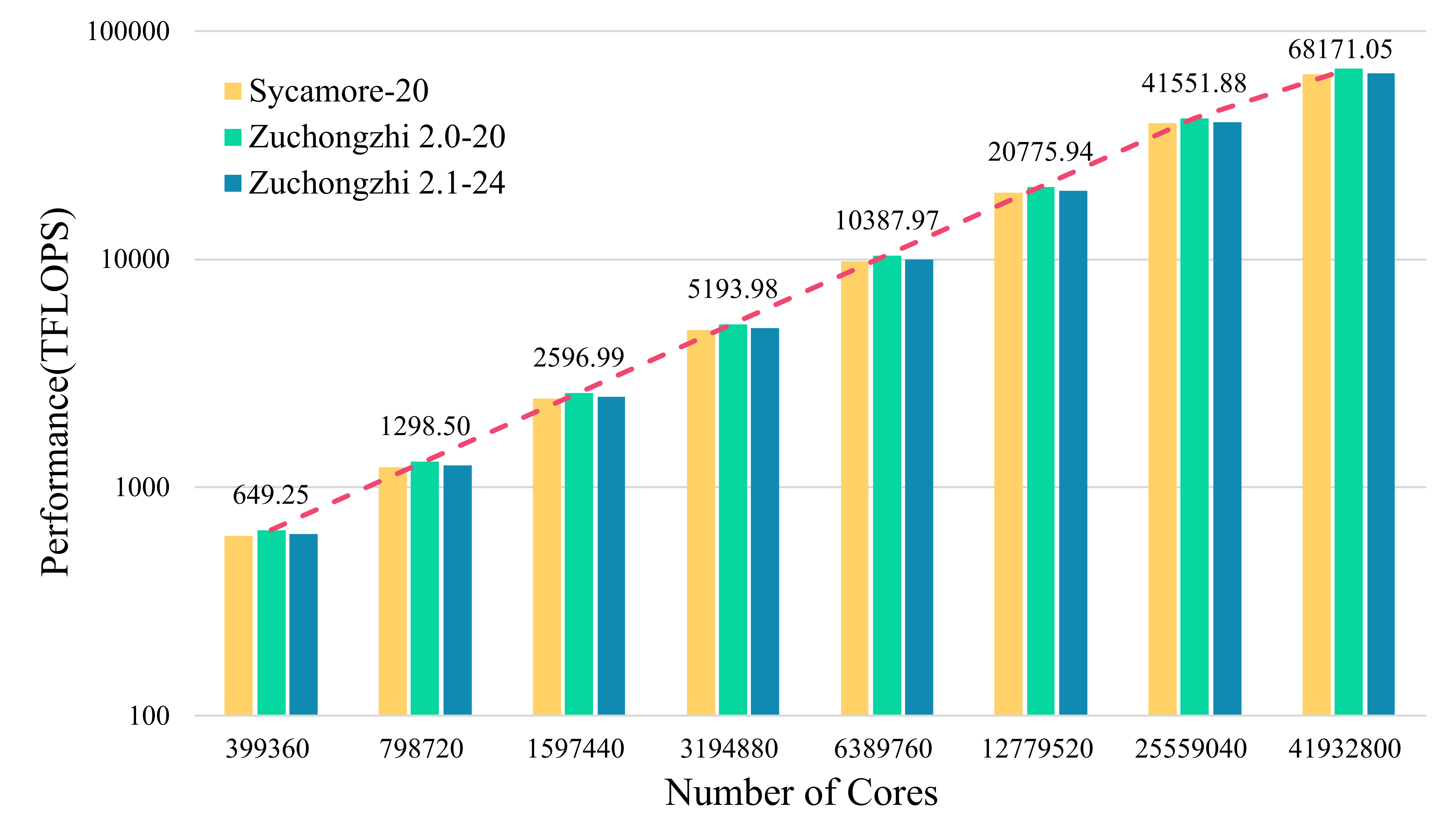}
\caption{Scaling of FLOPS performance against the number of cores used.
}
\label{fig:figS6}
\end{figure}

We further study the scaling of the FLOPS performance with against number of cores for contracting the tensor networks resulting from Sycamore-$20$, Zuchongzhi $2.0$-$20$ and Zuchongzhi $2.1$-$24$, with maximally allowed intermediate tensor sizes to be $2^{31}$, $2^{32}$ and $2^{32}$ respectively. The results are shown in Fig.~\ref{fig:figS6}. This numeric experiment explicitly demonstrates that strong scaling ($100\%$ parallelization efficiency) is achieved in all those cases.


\bibliographystyle{apsrev4-1}
\bibliography{refs}